# Structural and electronic signatures of strain-tunable marginally twisted bilayer graphene

Pei Ouyang[1,2], Jiawei Yu[1,2], Qian Li[1,2], Guihao Jia[1,2], Yuyang Wang[1,2], Kebin Xiao[1,2], Hongyun Zhang[1,2], Zhiqiang Hu[1,2], Pierre A. Pantaleón[3], Zhen Zhan[3, *], Shuyun Zhou[1,2], Francisco Guinea[3,4], Qi-Kun Xue[1,2,5,6,7, *], and Wei Li[1,2,7, *]

[1]*State Key Laboratory of Low-Dimensional Quantum Physics, Department of Physics, Tsinghua University, Beijing 100084, China*

[2]*Frontier Science Center for Quantum Information, Beijing 100084, China*

[3]*Imdea Nanoscience, Madrid 28015, Spain*

[4]*Donostia International Physics Center, San Sebastián 20018, Spain*

[5]*Beijing Academy of Quantum Information Sciences, Beijing 100193, China*

[6]*State Key Laboratory of Quantum Functional Materials and Department of Physics, Southern University of Science and Technology, Shenzhen 518055, China*

[7]*Hefei National Laboratory, Hefei 230088, China*

*Corresponding authors. E-mails: zhen.zhan@imdea.org;  qkxue@mail.tsinghua.edu.cn; weili83@tsinghua.edu.cn

**ABSTRACT**

Marginally twisted bilayer graphene having small twist angles is predicted to exhibit unique structural and electronic properties, though experimental characterization remains limited. Using scanning tunneling microscopy, we investigate such systems with twist angles of 0.06°–0.35°. AA-stacked regions reveal a pronounced tunneling spectral peak signifying highly localized electronic states. Conversely, AB domains display uniform multiple spectral peaks, indicative of strong lattice reconstruction and enhanced electronic homogeneity. We identify two distinct strain-induced domain walls: one exhibits a sharp $-120$ meV spectral peak (shear type), while the other shows distinct spectral characteristics (mixed shear-tensile type). Tight-binding calculations verify strain-driven transformations of both domain wall types and confirm direct observation of strain-mediated domain wall transitions. These results elucidate the electronic structure of marginally twisted bilayer graphene and establish strain as a control parameter for domain wall states.

**Keywords:** Twisted bilayer graphene, large moiré periods, domain wall, scanning tunneling microscopy

## INTRODUCTION

Twisted bilayer graphene (TBG) has attracted intense research interest due to its strongly correlated and topological states near the magic angle (~1.1°) [1–11]. In contrast, marginally twisted bilayer graphene (m-TBG) with twist angles far below 1° is predicted to host unique structural and electronic properties [2,12–18], yet has received relatively little experimental attention. At these small twist angles, strong lattice relaxation dramatically reconfigures the local atomic geometry. Specifically, this relaxation contracts AA regions (atoms vertically aligned), expands AB/BA regions (Bernal stacking) into triangular domains, and generates domain wall (DW) networks, creating geometries distinct from magic-angle TBG [12,16–18]. Such distinct behaviors in the moiré geometry pave a way for uniquely correlated and topological phenomena. For instance, a network of chiral 1D topological channels along the DWs is created due to the strong atomic and electronic reconstruction [17,19], robust proximity superconductivity in the quantum Hall regime is observed in the DWs [20], and phasons dominate the electron transport in m-TBG [21,22]. Consequently, direct local characterization of m-TBG's electronic structure remains essential.

DWs are typically classified as tensile or shear types based on their boundary orientation relative to the honeycomb lattice in TBG. Both types exhibit distinct properties that influence the system's electronic, magnetic, and optical characteristics [19,23–29]. In m-TBG, twist-induced lattice relaxation naturally generates networks of pure shear DWs hosting topological electronic states [15,16,20,30,31]. When large moiré periods (small twist angles) coexist with applied external bias, an energy gap opens in AB/BA domains while topologically protected helical edge modes emerge along DWs, enabling one-dimensional (1D) conduction [32–40]. Tensile DWs may appear in bilayer graphene under external strain [15]. Theoretical studies predict strain reshapes the triangular shear DW network and induces transitions to striped tensile DWs [41,42]. The atomic-scale dynamics of strain-driven DW transitions and their real-time observation remain significant unresolved questions.

In this study, we perform a systematic scanning tunneling microscopy (STM) investigation of m-TBG devices with twist angles ranging from 0.06° to 0.35°. The local density of states (LDOS) exhibits distinct signatures depending on the stacking configuration. A pronounced peak is observed at AA sites, reflecting strongly localized electronic states. Meanwhile, AB domains show spatially uniform multi-peak features indicative of enhanced electronic homogeneity arising from lattice reconstruction. Most significantly, we observe two types of DWs, distinguished in topography as the bright DW with enhanced contrast and the dark DW with reduced contrast. In tunneling spectra, dark DW shows a prominent peak at −120 meV, which is well reproduced by tight-binding calculations and attributed to a shear DW (DW-S). In contrast, bright DW lacks this resonance, and calculations identify it as a hybrid structure mixing shear and tensile DWs (DW-M). The specific definitions of DW-S and DW-M will be described later. These results confirm a strain-induced transition between the two DW types.

**RESULTS**

Our TBG device was fabricated on hexagonal boron nitride (hBN) [Fig. 1(a)]. STM topography [Fig. 1(b)] reveals distorted moiré triangles exhibiting strong lattice reconstruction and strain [24,43]. Specifically, AA regions shrink and connect via quasi-1D DWs of varying lengths. Half the moiré triangles display faint or near-threshold DWs, appearing weakly visible or near the detection limit. Accordingly, two DW types are identified through topographic contrast in Fig. 1(b): DW-M (bright stripes) and DW-S (extremely low contrast, exemplified by the yellow-dashed specimen), indicating distinct electronic properties – with schematics in Fig. 1(c). The lower panel of Fig. 1(c) defines tensile and shear DWs exhibiting distinct electronic structures [28] (Supplemental Material Sec. 1), which we later demonstrate directly correspond to the two observed DW types.

The twist angle and strain tensor can be obtained once the primitive lattice periods of each moiré triangle are determined (see details in Sec. 2 of SM). The twist angles across the sample [Fig. 1(e) and upper panel of (g)] gradually vary from 0.06° to 0.35°. Crucially, within each moiré triangle, both uniaxial strain [$\epsilon_{xx}$ = 0.06-0.9%, Fig. 1(f) and lower panel of (g)] and shear strain (Fig. S4) are calculated, providing a platform to investigate the combined effects of twist-angle and strain in m-TBG.

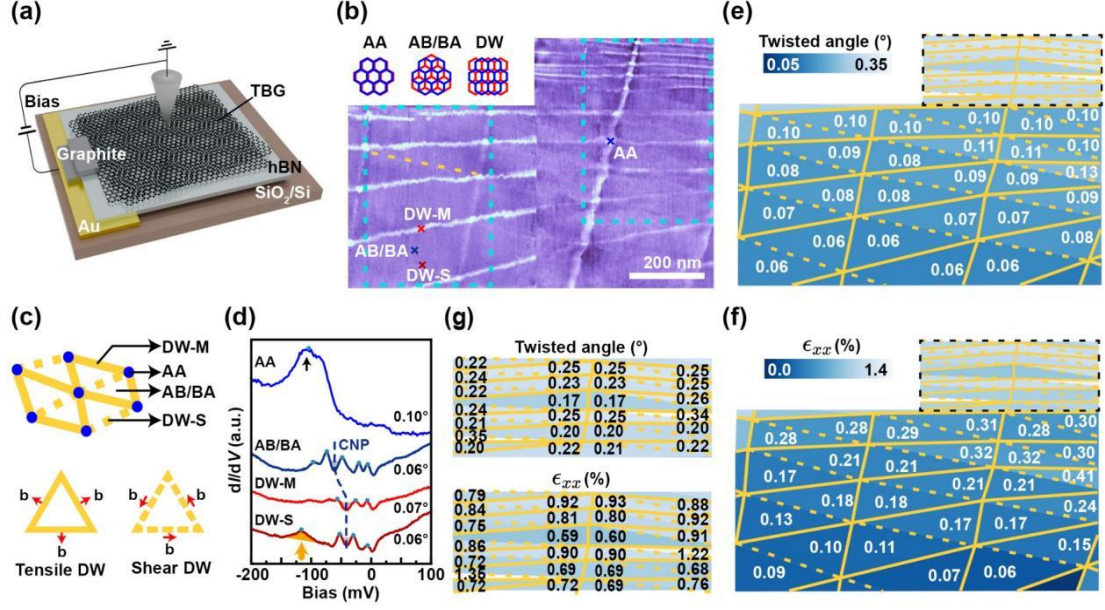

**Figure 1. Topography and tunneling spectra at different stacking regions of a marginally twisted bilayer graphene device.** (**a**) Sample configuration. TBG is transferred onto hBN substrate and bias voltage $V_b$ between the STM tip and TBG is applied through a graphite electrode. (**b**) Large-area STM topography of marginally twisted bilayer graphene (m-TBG) device spliced by three images. Left: 500 nm × 500 nm, bias voltage $V_b$ = −2000 mV, tunneling current $I_t$ = 20 pA; upper right: 500 nm × 500 nm, $V_b$ = −1000 mV, $I_t$ = 20 pA; right: 760 nm × 500 nm, $V_b$ = −2000 mV, $I_t$ = 20 pA. Three types of stacking configurations (AA, AB/BA, and shear DW) are shown on the upper left panel. Colored crosses mark spectroscopy sites in (d): AA (blue), AB/BA (dark blue), DW-M (red), DW-S (dark red). The left cyan dashed square marks the region in Fig. 3(a). The right cyan dashed square marks the region in Fig. 3(b). Yellow dashed line marks an individual DW-S. (**c**) Schematic of moiré pattern across different stacking regions (AA, AB/BA) and domain wall types (DW-M and DW-S). Symbolic correspondence: AA sites (blue dots), DW-S (yellow dashed lines), and DW-M (yellow solid lines). The lower panel shows a schematic of a tensile DW (Burger vector **b** perpendicular to the DW boundary) and a shear DW (**b** parallel to the DW boundary). (**d**) d$I$/d$V$ spectra on four sites in (b). The dark blue dashed line indicates the shifted charge neutrality point (CNP). The black arrow marks the peak position of the AA site. The orange arrow marks the peak position of DW-S. (**e**) Spatial distribution of twist angles in TBG sample, corresponding to the topography shown in (b). DW-S is highlighted by a dashed yellow line, and DW-M is highlighted by solid yellow line. (**f**) Spatial distribution of uniaxial strain $\epsilon_{xx}$ in TBG sample. (**g**) Zoomed views of black dashed regions in (e) (upper panel), and (f) (lower panel). Set point: (d) $V_b$ = −200 mV, $I_t$ = 200 pA.

Tunneling spectroscopy measurements reveal distinct electronic signatures associated with different stacking configurations [Fig. 1(d)]. The AA sites show a pronounced peak-like feature near

−100 meV, attributed to lattice relaxation and consistent with previous reports of reconstructed m-TBG [44]. Notably, the AB/BA stacked regions display several intrinsic sharp peaks in the d$I$/d$V$ spectra. The spectrum measured at the AB region shows a dip-like feature at −60 meV. From AA peak and AB/BA dip positions, we estimate the charge neutrality point (CNP) is around −60 meV. The shifted CNP (marked by blue dashed line) in AA and AB/BA regions indicates different local electron doping, consistent with previous measurement on hBN-supported graphene [45]. Domain walls show markedly different LDOS (detailed in Fig. 2), with only DW-S exhibiting a pronounced −120 meV peak. In contrast to magic-angle TBG, m-TBG lacks prominent "remote band" signatures, and its flat band peak broadens markedly at AA sites (Fig. S5).

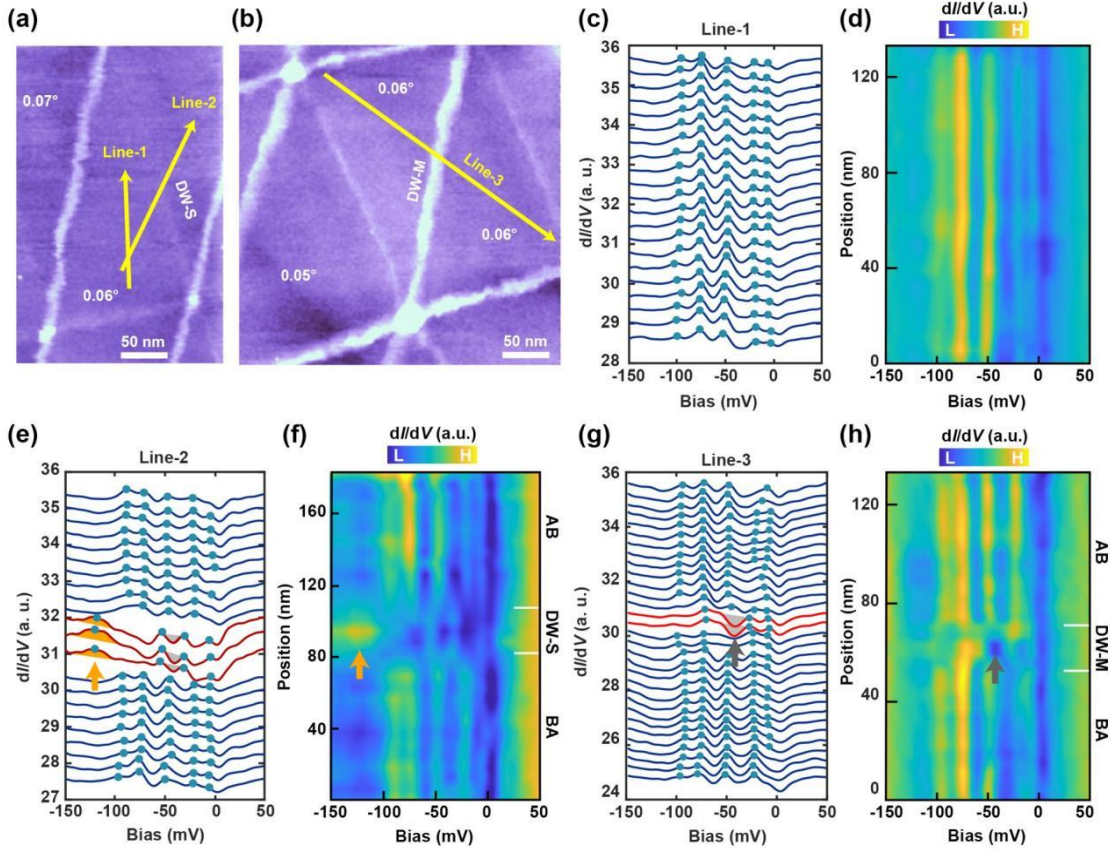

**Figure 2. Tunneling spectra of AB/BA domains and DWs in m-TBG.** (**a**) STM topography (220 nm × 350 nm, $V_b$ = −2000 mV, $I_t$ = 20 pA) of m-TBG. (**b**) STM topography (350 nm × 350 nm, $V_b$ = −2000 mV, $I_t$ = 20 pA) of another m-TBG region. (**c**) d$I$/d$V$ spectra taken along Line-1 in (a), demonstrating spectral uniformity within an AB domain. Each curve is vertically offset for clarity. (**d**) Colormap of (**c**) further visualizes the invariance of peak positions. (**e**) d$I$/d$V$ spectra taken along Line-2 in (a), with dark red curves highlighting DW-S regions with a peak at -120 meV (orange arrow). (**f**) Colormap of (e), highlighting the peak at -120 meV (orange arrow) in DW-S. (**g**) d$I$/d$V$ spectra taken along Line-3 in (b), where red curves indicate DW-M with spectra showing a dip at -40 meV (grey arrow). (**h**) Colormap of (g), highlighting the dip at -40 meV in DW-M. Green dots in (c), (e) and (g) mark the positions of spectral peaks. Set point: (c-h) $V_b$ = -200 mV, $I_t$ = 200 pA.

To further investigate the electronic structure of m-TBG, we performed detailed tunneling spectroscopy measurements across different stacking configurations. Figure 2(a) displays a STM topographic image where the yellow line (Line-1) traces a path along an AB-stacked region within a moiré triangle. The corresponding spectra [Fig. 2(c)] reveal five well-defined peaks in the energy range of 0 to -100 meV, with remarkably uniform energy spacing. The spatially resolved d$I$/d$V$ colormap along this trajectory [Fig. 2(d)] highlights the exceptional electronic uniformity within the AB/BA regions, confirming the homogeneity of the reconstructed states. These spectral features are attributed to the intrinsic electronic structure of m-TBG, arising from strong lattice reconstruction in the large-period moiré superlattice under strain [15,44,46]. Alternative explanation for these peaks, such as electron confinement, is excluded based on analysis of the size of moiré superlattice (details in Sec. 1 of SM).

As mentioned before, STM topography clearly shows two types of DWs: DW-M with enhanced topographic contrast and DW-S with reduced contrast. Spectroscopic linecuts across DWs and AB/BA regions [traced by Line-2 in Fig. 2(a) and Line-3 in Fig. 2(b)] reveal their distinct electronic signatures. The spectra measured across DW-S [Fig. 2(e)] exhibit a pronounced peak at -120 meV, along with a noticeable shift of the CNP from -60 meV in the AB/BA regions to -40 meV at the DW. The corresponding d$I$/d$V$ colormap [Fig. 2(f)] captures the spatial evolution of these features across the BA-(DW-S)-AB transition, highlighting the modification of electronic structure at the DW-S.

Spectra across the DW-M (traced by Line-3) reveals a characteristic dip at -40 meV [Fig. 2(g)], consistent with the CNP shift observed in DW-S. Notably, the -120 meV peak is absent on the DW-M. The corresponding d$I$/d$V$ colormap [Fig. 2(h)] captures the evolution of the electronic states across the BA-(DW-M)-AB transition, offering a detailed view of the DW-M electronic structure (Fig. S6-S7). Despite showing similar CNP shifts, the two types of DWs exhibit distinct spectroscopic features, indicative of fundamentally different electronic origins [23,26,28]. Within a single moiré triangle, one edge forms a DW-S, and the remaining two edges are DW-Ms subject to different strain conditions (see also Fig. S8-S11). Here, the pseudo-magnetic field origin for the observed spectra on DWs is excluded through an analysis of spatial width of DWs (see Sec. 1 of SM).

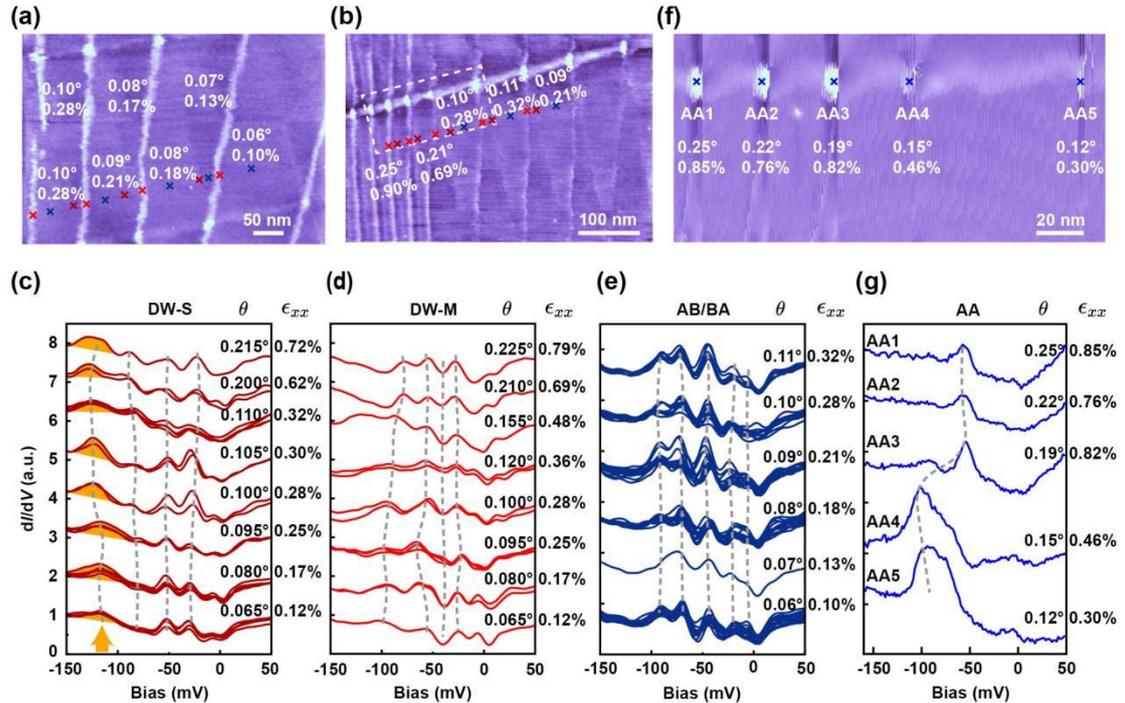

**Figure 3**. **Tunneling spectra of DWs, AB regions, and AA sites in m-TBG with different twist angles and strain $\epsilon_{xx}$.** (**a**) STM topography (500 nm × 350 nm, $V_b$ = -2000 mV, $I_t$ = 20 pA) of m-TBG with twist angles ranging from 0.06° to 0.10° (with $\epsilon_{xx}$ ranging from 0.10% to 0.28%), taken from the cyan dashed square on the left in Fig. 1(b). (**b**) STM topography (500 nm × 350 nm, $V_b$ = -1000 mV, $I_t$ = 20 pA) of TBG region with twist angles ranging from 0.09° to 0.25° (with $\epsilon_{xx}$ ranging from 0.21% to 0.90%), taken from the upper right cyan square in Fig. 1(b). (**c**)-(**e**) Twist angle evolution of electronic states across different DW configurations: (**c**) d$I$/d$V$ spectra of DW-S (dark red crosses) from (a) and (b), highlighting the peak at -120 meV (orange arrow); (**d**) d$I$/d$V$ spectra of DW-M (red crosses) from (a) and (b); (**e**) d$I$/d$V$ spectra from AB/BA domains (dark blue crosses) in (a) and (b), demonstrating spectral uniformity across AB/BA domains. (**f**) Zoomed-in STM topography (180 nm × 90 nm, $V_b$ = -200 mV, $I_t$ = 20 pA) of the area marked by the white dashed square in (b). (**g**) Corresponding d$I$/d$V$ spectra acquired at AA sites with different twist angles (vertical offsets applied for clarity). Set point for (c-e) and (g): $V_b$ = -200 mV, $I_t$ = 200 pA.

Figures 3(a) and 3(b) show STM topographies corresponding to the areas within cyan dashed squares of Fig. 1(b), covering a twist angle range between 0.065° and 0.225°. Remarkably, the spectroscopic features demonstrate exceptional robustness against twist-angle variations. For DW-S, the characteristic −120 meV peak persists with only minimal energy variation (±5 mV), consistently accompanied by a sharp dip-like feature near -40 meV. The AB/BA regions exhibit striking spectral uniformity, with all five characteristic peaks between 0 and -100 meV remaining essentially unchanged for twist angles between 0.06° and 0.11°. In contrast, the electronic structures in AA sites are much more sensitive to the twist angles and strain. Figure 3(f) shows an area including five AA sites with different twist angles. Their characteristic LDOS peaks shift gradually from -110 mV to -50 mV with increasing twist angle, indicating a continuous modification of the local electronic structure in response to changes in the moiré pattern and twist angle. The strain-dependent trend is qualitatively consistent with the evolution observed as a function of twist angle.

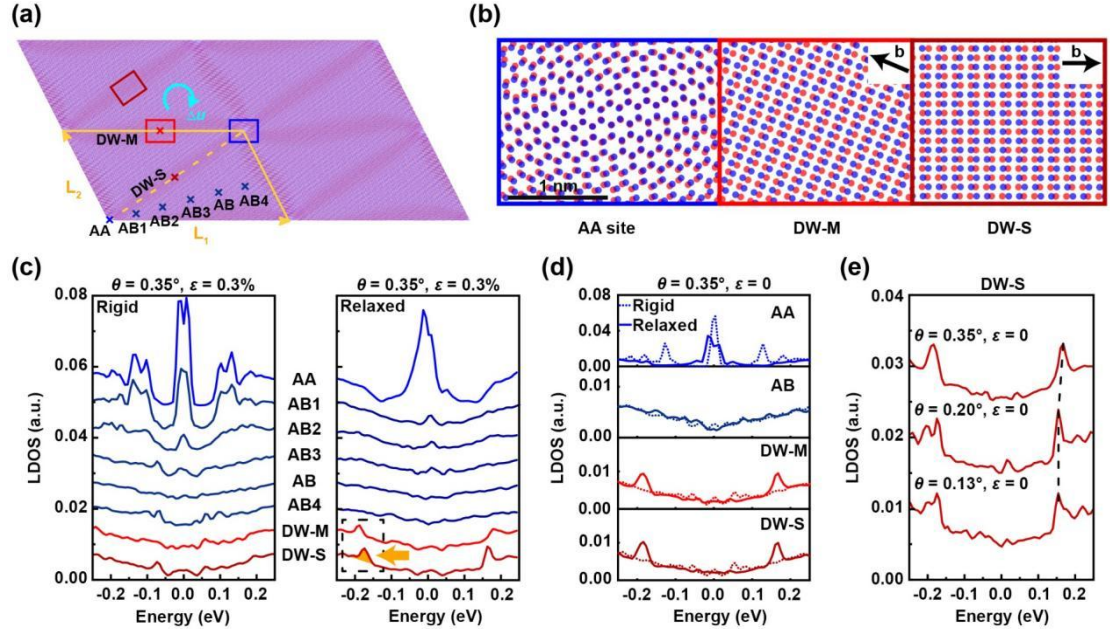

**Figure 4. Theoretical properties of TBG with marginal twist angles and strain.** (**a**) The 2×2 moiré supercell of relaxed TBG with a twist angle $\theta$ = 0.35° and a uniaxial hetero-strain $\varepsilon$ = 0.3%.

The blue, red, and dark red rectangles mark the AA region and two DW regions, referred to as DW-M and DW-S. $L_1$ and $L_2$ denote the moiré lattice vectors. Cyan arrow denotes the atom displacement $\Delta \boldsymbol{u}$. (**b**) Atomic structures of the AA site, DW-M, and DW-S in the relaxed TBG. Black arrows denote the Burger vector **b**. (**c**) Calculated LDOS variations across different positions in TBG with and without (rigid) lattice relaxation. The corresponding measurement points are illustrated in (a). The DW-related peaks are highlighted by a black dashed rectangle. The orange arrow highlights the peak of DW-S. (**d**) LDOS at various positions in TBG with the same twist angle but without strain. The dashed line and solid line correspond to the rigid and relaxed structures, respectively. (**e**) Twist-angle-dependent evolution of the LDOS at DW-S in unstrained TBG.

We perform tight-binding (TB) calculations (details in Sec. VI in SM) to investigate the electronic structure of m-TBG and to understand key experimental observations, such as the robust d$I$/d$V$ spectra in the AB/BA regions and the distinct features at the DWs. The full atomistic TB model is an accurate and powerful tool to investigate strain and lattice relaxation effects on local properties of moiré systems. In practice, the moiré patterns in our sample span hundreds of nanometers, which are beyond the computational limits of atomistic simulations. Significantly, the relevant factors to influence the electronic properties of our sample are lattice relaxation and strain. Therefore, we constructed a m-TBG model with a relatively small twist angle of 0.35° and introduced uniaxial hetero-strain. The relaxed structure exhibits a DW network [Fig. 4(a)] with two distinct atomic configurations [Fig. 4(b), S8-S9]: DW-S (dark red rectangle) and DW-M (red rectangle), corresponding to experimental counterparts as established below. Local atomic structure analysis reveals the angle $β$ (between domain boundary and Burger vector) is 90° for DW-S (pure shear type) versus < 90° for DW-M (mixed shear-tensile type) (Fig. S10). The LDOS of both long and short DW-Ms within a single moiré triangle are also presented in Fig. S10. Despite different local strain environments, d$I$/d$V$ spectra of both long and short DW-M (Fig. S11) confirm their electronic similarity. Strain drives this structural transition and distorts the moiré primitive cell from hexagonal to quasi-rectangle, matching experimental observations. Lattice relaxation further shrinks AA regions and expands AB domains into triangles while inducing DWs along the boundaries [Fig. 4(b), Figs. S7, S8].

LDOS calculations (proportional to d$I$/d$V$ spectra) show enhanced uniformity across expanded AB regions post-relaxation [Fig. 4(c)], consistent with experimental homogeneity. Specifically, the LDOS in the AB1, AB2, AB3 and AB positions of the relaxed case behave similarly, whereas they show an obvious difference across the AB region in the rigid case. Discrepancies between experimental and theoretical LDOS in AB regions likely arise from twist angle and strain variation. Crucially, under strain, DW-S exhibits a pronounced -180 meV peak while DW-M shows only weak low-energy resonances [the features within the dashed rectangle in Fig. 4(c)], contrasting sharply with unstrained calculations where both DW types display identical -180 meV peaks [Fig. 4(d)].

This establishes direct correspondence: (1) Experimental DW-S (low-contrast topography, -120 meV peak) matches theoretical shear-type DW (-180 meV peak); (2) Experimental DW-M (bright topography, no peak) matches theoretical mixed-type DW (no peak). The energy discrepancy (-180 meV vs -120 meV) arises from a combination of doping effects, twist angle variation and strain difference [45, 47]. Strain eliminates the characteristic resonance by driving shear-to-mixed transitions, confirming strain-mediated DW reconfiguration [41,42]. Our simulations thus qualitatively reproduce all key spectroscopic features through synergistic strain and relaxation effects.

**CONCLUSION**

Using scanning tunneling microscopy and tight-binding modeling, we demonstrate deterministic strain control of domain wall (DW) states in marginally twisted bilayer graphene (0.06°–0.35°). Our key findings include: (i) Systematic local characterization of the electronic properties of m-TBG, which behave significantly differently from those in the magic angle TBG, (ii) Discovery of a strain-sensitive -120 meV electronic resonance exclusive to shear-type DWs (DW-S), (iii) Strain-driven transformation to mixed shear-tensile DWs (DW-M) with distinct electronic signatures, directly visualized through atomic-scale spectroscopy.

This work provides fundamental insights into the intrinsic electronic properties of large-period moiré superlattices in TBG, with particular emphasis on how lattice reconstruction and strain govern these electronic structures. More importantly, we demonstrate experimental evidence of strain-driven phase transitions between the two types of domain walls. Our findings establish strain as a key tuning parameter for engineering 1D domain walls, thereby enabling manipulation of electronic, transport, and optical properties in marginally twisted TBG.

**METHODS**

**STM measurements.**

Our experiments were performed using a commercial ultrahigh-vacuum scanning tunneling microscope (USM-1200, Unisoku) maintained at a base temperature of 4.2 K, equipped with optical microscopy capabilities. The system maintained a base pressure of $1.0 \times 10^{-10}$ Torr. The TBG device was degassed at 170 °C under ultrahigh vacuum before being transferred into the STM. Scanning tunneling spectroscopy (STS) measurements were conducted using a standard lock-in amplification technique with a modulation frequency of 973.0 Hz. Differential conductance ($dI/dV$) spectra were acquired by disabling the feedback loop while ramping the DC bias voltage.

**TBG device**.

Twisted bilayer graphene (tBLG) samples were fabricated using a clean, dry transfer technique. Monolayer graphene was first exfoliated onto a clean $SiO_2$/Si substrate. A hexagonal boron nitride (hBN) flake mounted on a polydimethylsiloxane (PDMS) stamp was aligned above the graphene under an optical microscope. One half of the graphene flake was picked up by the hBN, forming a PDMS/hBN/graphene stack. The remaining half of the graphene was then rotated to the desired twist angle and subsequently picked up to complete the twisted bilayer structure. The resulting PDMS/hBN/tBLG stack was flipped, and the hBN/tBLG was transferred onto a gold-coated substrate using a second PDMS stamp. The structural cleanliness and integrity of the tBLG were crucial for acquiring high-quality STM data. To ensure good electrical contact during STM measurements, a graphite flake was used to bridge the tBLG and the gold substrate.

**SUPPLEMENTARY DATA**

Supplementary data are available at NSR online.

ACKNOWLEDGEMENTS

We thank Biao Lian, Qunyang Li, Gerardo G. Naumis, and Federico Escudero for helpful discussions.

FUNDING

The experimental work was supported by the National Natural Science Foundation (92365201, 52388201 and 11427903), the Ministry of Science and Technology of the People's Republic of China (2022YFA1403100), the Innovation Program for Quantum Science and Technology (2021ZD0302402), and the Beijing Advanced Innovation Center for Future Chip (ICFC). Z.Z. acknowledges support funding from the European Union's Horizon 2020 research and innovation programme under the Marie Skłodowska-Curie grant agreement (101034431), and from the "Severo Ocho" Programme for Centres of Excellence in R\&D (CEX2020-001039-S / AEI /10.13039/501100011033). P.A.P and F.G. acknowledge funding from Comunidad de Madrid, Spain (NMAT2D), SprQuMat and (MAD2D-CM)-MRR MATERIALES AVANZADOS-IMDEA-NC. P.A.P acknowledges funding of the Julian Schwinger Foundation for Physics Research (JSF-24-05-0002).

AUTHOR CONTRIBUTIONS

W.L., S.Z., and Q.-K.X. conceived and supervised the research project. J.Y., K.X., and Z.H. performed the STM experiments. Q.L., H.Z., and S.Z. prepared the TBG device. Z.Z. and F.G. performed the calculations. W.L., J.Y., P.O., Z.Z., G.J., Y.W., P.A.P., and F.G. analyzed the data. W.L., P.O., Z.Z., and G.J. wrote the manuscript with input from all other authors.

**Conflict of interest statement.** None declared.

REFERENCES

1. Bistritzer R, MacDonald A H. Moiré bands in twisted double-layer graphene. *Proc. Natl. Acad. Sci.* 2011; **108**: 12233-12237.
2. Nam N N T, Koshino M. Lattice relaxation and energy band modulation in twisted bilayer graphene. *Phys. Rev. B* 2017; **96**: 075311.
3. Andrei E Y, MacDonald A H. Graphene bilayers with a twist. *Nat. Mater.* 2020; **19**: 1265-1275.
4. Cao Y, Fatemi V, Demir A, *et al.* Correlated insulator behaviour at half-filling in magic-angle graphene superlattices. *Nature* 2018; **556**: 80-84.
5. Xie Y, Lian B, Jäck B, *et al.* Spectroscopic signatures of many-body correlations in magic-angle twisted bilayer graphene. *Nature* 2019; **572**: 101-105.
6. Wong D, Nuckolls K P, Oh M, *et al.* Cascade of electronic transitions in magic-angle twisted bilayer graphene. *Nature* 2020; **582**: 198-202.
7. Kerelsky A, McGilly L J, Kennes D M, *et al.* Maximized electron interactions at the magic angle in twisted bilayer graphene. *Nature* 2019; **572**: 95-100.
8. Jiang Y, Lai X, Watanabe K, *et al.* Charge order and broken rotational symmetry in magic-

angle twisted bilayer graphene. *Nature* 2019; **573**: 91-95.

9. Choi Y, Kemmer J, Peng Y, *et al.* Electronic correlations in twisted bilayer graphene near the magic angle. *Nat. Phys.* 2019; **15**: 1174-1180.
10. Cao Y, Fatemi V, Fang S, *et al.* Unconventional superconductivity in magic-angle graphene superlattices. *Nature* 2018; **556**: 43-50.
11. Zhang Z, Yang J, Xie B, *et al.* Cascade of Zero-field Chern insulators in Magic-angle bilayer graphene. *Natl. Sci. Rev.* 2025: nwaf265.
12. Wijk M M van, Schuring A, Katsnelson M I, *et al.* Relaxation of moiré patterns for slightly misaligned identical lattices: graphene on graphite. *2D Mater.* 2015; **2**: 034010.
13. Guinea F, Walet N R. Continuum models for twisted bilayer graphene: Effect of lattice deformation and hopping parameters. *Phys. Rev. B* 2019; **99**: 205134.
14. Walet N R, Guinea F. The emergence of one-dimensional channels in marginal-angle twisted bilayer graphene. *2D Mater.* 2019; **7**: 015023.
15. Mesple F, Walet N R, Trambly de Laissardière G, *et al.* Giant atomic swirl in graphene bilayers with biaxial heterostrain. *Adv. Mater.* 2023; **35**: 2306312.
16. Liu Y W, Su Y, Zhou X F, *et al.* Tunable lattice reconstruction, triangular network of chiral One-dimensional states, and bandwidth of flat bands in magic angle twisted bilayer graphene. *Phys. Rev. Lett.* 2020; **125**: 236102.
17. Yoo H, Engelke R, Carr S, *et al.* Atomic and electronic reconstruction at the van der Waals interface in twisted bilayer graphene. *Nat. Mater.* 2019; **18**: 448-453.
18. Gargiulo F, Yazyev O V. Structural and electronic transformation in low-angle twisted bilayer graphene. *2D Mater.* 2017; **5**: 015019.
19. Ju L, Shi Z, Nair N, *et al.* Topological valley transport at bilayer graphene domain walls. *Nature* 2015; **520**: 650-655.
20. Barrier J, Kim M, Kumar R K, *et al.* One-dimensional proximity superconductivity in the quantum Hall regime. *Nature* 2024; **628**: 741-745.
21. Ochoa H. Moiré-pattern fluctuations and electron-phason coupling in twisted bilayer graphene. *Phys. Rev. B* 2019; **100**: 155426.
22. Boschi A, Ramos-Alonso A, Mišeikis V, *et al.* Phason-driven temperature-dependent transport in moiré graphene. 2025, DOI: 10.48550/arXiv.2511.01691.
23. Koshino M. Electronic transmission through AB-BA domain boundary in bilayer graphene. *Phys. Rev. B* 2013; **88**.
24. Alden J S, Tsen A W, Huang P Y, *et al.* Strain solitons and topological defects in bilayer graphene. *Proc. Natl. Acad. Sci.* 2013; **110**: 11256-11260.
25. Lin J, Fang W, Zhou W, *et al.* AC/AB Stacking Boundaries in Bilayer Graphene. *Nano Lett.* 2013; **13**: 3262-3268.
26. San-Jose P, Gorbachev R V, Geim A K, *et al.* Stacking Boundaries and Transport in Bilayer Graphene. *Nano Lett.* 2014; **14**: 2052-2057.
27. González J. Confining and repulsive potentials from effective non-Abelian gauge fields in graphene bilayers. *Phys. Rev. B* 2016; **94**: 165401.
28. Timmel A, Mele E J. Dirac-Harper Theory for One-Dimensional Moiré Superlattices. *Phys. Rev. Lett.* 2020; **125**: 166803.
29. Jiang L, Wang S, Zhao S, *et al.* Soliton-Dependent Electronic Transport across Bilayer Graphene Domain Wall. *Nano Lett.* 2020; **20**: 5936-5942.

30. Mahapatra P S, Garg M, Ghawri B, *et al.* Quantum Hall Interferometry in Triangular Domains of Marginally Twisted Bilayer Graphene. *Nano Lett.* 2022; **22**: 5708-5714.
31. Wang H C, Hsu C H. Electrically tunable correlated domain wall network in twisted bilayer graphene. *2D Mater.* 2024; **11**: 035007.
32. Anđelković M, Covaci L, Peeters F M. DC conductivity of twisted bilayer graphene: Angle-dependent transport properties and effects of disorder. *Phys. Rev. Mater.* 2018; **2**: 034004.
33. Zhang F, MacDonald A H, Mele E J. Valley Chern numbers and boundary modes in gapped bilayer graphene. *Proc. Natl. Acad. Sci.* 2013; **110**: 10546-10551.
34. San-Jose P, Prada E. Helical networks in twisted bilayer graphene under interlayer bias. *Phys. Rev. B* 2013; **88**: 121408.
35. Huang S, Kim K, Efimkin D K, *et al.* Topologically Protected Helical States in Minimally Twisted Bilayer Graphene. *Phys. Rev. Lett.* 2018; **121**: 037702.
36. Zheng Q, Hao C Y, Zhou X F, *et al.* Tunable Sample-Wide Electronic Kagome Lattice in Low-Angle Twisted Bilayer Graphene. *Phys. Rev. Lett.* 2022; **129**: 076803.
37. Vaezi A, Liang Y, Ngai D H, *et al.* Topological Edge States at a Tilt Boundary in Gated Multilayer Graphene. *Phys. Rev. X* 2013; **3**: 021018.
38. Rickhaus P, Wallbank J, Slizovskiy S, *et al.* Transport Through a Network of Topological Channels in Twisted Bilayer Graphene. *Nano Lett.* 2018; **18**: 6725-6730.
39. Hou Z, Yuan K, Jiang H. Arrays of one-dimensional conducting channels in minimally twisted bilayer graphene. *Phys. Rev. B* 2024; **110**: L161406.
40. De Beule C, Dominguez F, Recher P. Network model and four-terminal transport in minimally twisted bilayer graphene. *Phys. Rev. B* 2021; **104**: 195410.
41. Lebedeva I V, Popov A M. Energetics and Structure of Domain Wall Networks in Minimally Twisted Bilayer Graphene under Strain. *J. Phys. Chem. C* 2020; **124**: 2120-2130.
42. Lebedeva I V, Popov A M. Two Phases with Different Domain Wall Networks and a Reentrant Phase Transition in Bilayer Graphene under Strain. *Phys. Rev. Lett.* 2020; **124**: 116101.
43. Sinner A, Pantaleón P A, Guinea F. Strain-Induced Quasi-1D Channels in Twisted Moir\'e Lattices. *Phys. Rev. Lett.* 2023; **131**: 166402.
44. Trambly de Laissardière G, Mayou D, Magaud L. Numerical studies of confined states in rotated bilayers of graphene. *Phys. Rev. B* 2012; **86**: 125413.
45. Yu J, Jia G, Li Q, *et al.* Twist angle driven electronic structure evolution of twisted bilayer graphene. 2024, DOI: 10.48550/arXiv.2406.20040.
46. Huder L, Artaud A, Le Quang T, *et al.* Electronic Spectrum of Twisted Graphene Layers under Heterostrain. *Phys. Rev. Lett.* 2018; **120**: 156405.
47. Choi Y, Kim H, Lewandowski C, *et al.* Interaction-driven band flattening and correlated phases in twisted bilayer graphene. *Nat. Phys.* 2021; **17**: 1375-1381.